\documentclass[10pt,draft]{dis03}
\usepackage{epsf,amsmath,rotating,axodraw}

\newcommand{\gsim}{\raisebox{-0.13cm}{~\shortstack{$>$ \\[-0.07cm] $\sim$}}~}

\newcommand{\beq}{\begin{equation}}
\newcommand{\eeq}{\end{equation}}
\newcommand{\beqn}{\begin{eqnarray}}
\newcommand{\eeqn}{\end{eqnarray}}

\begin{document}

\title{QCD CORRECTIONS TO PHOTOPRODUCTION\\
OF $W$ BOSONS AT HERA
\thanks{Presented at DIS2003, April 23 -- 27, St. Petersburg, Russia}}

\author{C.~SCHWANENBERGER  \\
DESY \\
Notkestr. 85, D-22607 Hamburg, Germany\\
E-mail: schwanen@mail.desy.de }

\maketitle

\begin{abstract}
\noindent $W$ bosons can be produced in the channels $e^\pm p\to W^\pm + X$ at
HERA thus allowing anomalous trilinear couplings among the
gauge bosons to be probed.  We discuss the next-to-leading order (NLO) QCD
corrections to the photoproduction of $W$ bosons
with finite transverse momentum at HERA. The higher-order QCD corrections
reduce the factorization scale dependence significantly and modify the
leading-order (LO) cross sections by $\pm {\cal O}(10\%)$.
\end{abstract}

\section{Introduction}
The study of $W$ bosons at different colliders serves as an important test
of the Standard Model and its possible extensions. $W$ bosons can be
produced at the $ep$ collider HERA with a center of mass (c.m.)
energy $\sqrt{s}\approx 318$ GeV which is achieved by colliding
electrons/positrons with energy $E_e=27.5$ GeV and protons with energy
$E_p=920$ GeV. Since the production cross sections for the
processes $e^\pm p \to e^\pm W + X$ reach values of about 1 $pb$ at HERA,
the production mechanisms of $W$ bosons can be studied and the existence
of anomalous $WW\gamma$ trilinear couplings can be probed
\cite{heratri,dubinin}. Moreover, $W$ boson
production represents an important SM background to several new physics
searches. In particular it is the dominant SM process leading to isolated
high energy lepton events with missing transverse momentum
\cite{heratri,muon}. In order to determine potential discrepancies between
measurements and Standard Model (SM) predictions, the latter have to be
sufficiently accurate and reliable. This is not guaranteed for the
available LO calculations of $W$ boson production
(see Ref.~\cite{baur,paper} and references therein). For an unambiguous test of
anomalous contributions, 
it is necessary to extend the previous analyses to NLO accuracy. The first
step in this direction has been made in Ref.~\cite{NRS}, where the QCD
corrections to the total resolved photoproduction cross section have been
determined. However, the result cannot be used for $W$ boson production
with large transverse momentum which is dominated
by direct photoproduction.
In the following we shall describe the QCD corrections to direct 
photoproduction which were published in Ref.~\cite{paper}. 
The procedure of implementing the QCD
corrections in Monte Carlo programs by a reweighting method is described in
Ref.~\cite{reweight}.

\section{Leading Order}
$W$ boson production at $ep$ colliders is mediated by photon, $Z$
and $W$ exchange between the electron/positron and the hadronic currents of
the process. In general two phase-space regions are distinguished:
the deep inelastic (DIS) regime at
large $Q^2$ and the photoproduction regime at small $Q^2$, $Q^2$ being the
negative square of the transferred momentum from the electron/positron. 
Details of the calculation of the
photoproduction cross section are given in Ref.~\cite{paper}.

The leading direct photon process $\gamma q\to q'W$ 
develops a singularity at LO when the final state quark $q'$ becomes collinear
with the initial state photon. However, the finite transverse momentum
$p_{TW}$ of the $W$ boson has to be balanced by the final state quark, so
that this singularity does not occur at LO for non-vanishing
$p_{TW}$. The small $Q^2$ region includes the contribution of
the hadronic component of the photon giving rise to $W+jet$ production
via $q\bar q'\to Wg$ and the crossed processes $gq(\bar q) \to W q'(\bar q')$.
The treatment of the DIS region is straightforward.

The direct, resolved and DIS contributions add up to the total
$p_{TW}$ distribution. Direct photoproduction forms the dominant
contribution, while the DIS part is smaller but significant. The
resolved component is negligible for $p_{TW} \gsim 15$ GeV \cite{spira}.
The dependence on the specific value of the cut $Q^2_{max}$ which
separates the DIS and photoproduction regimes is below the per-cent level
\cite{paper} and thus sufficiently small. 

\section{QCD Corrections}
For the dominant direct part we have evaluated the NLO QCD corrections.
They consist of two parts, the
virtual and real corrections. The virtual corrections are built up by all
one-loop diagrams which are generated by virtual gluon exchange. They have
been computed via dimensional regularization in $n=4-2\epsilon$
dimensions. The quarks have been treated as massless particles. 

The real corrections which
originate from gluon radiation off the quark lines and the corresponding
crossed contributions with the gluon in the initial state
have been calculated by means of the massless dipole
subtraction
method introduced in Ref.~\cite{catsey}. 
They cancel the infrared and collinear
divergences of the virtual corrections and the collinear
singularity of the counter term due to
the renormalization of the parton densities at NLO, respectively. The
NLO parton densities have been defined in the $\overline{\rm MS}$ scheme.
This procedure allows to calculate the real matrix elements in 4 dimensions.

\section{Results}
We analyzed our final results for direct photoproduction of $W$ bosons
plus one jet as well as the inclusive process $W+X$, i.e.~without
defining jets.
Details of the jet requirements, our choice of parton density functions and 
values of the relevant numerical parameters can be found in Ref.~\cite{paper}.

\begin{figure}[!hbtp]
\vspace*{0.5cm}

\begin{turn}{-90}%
\epsfxsize=8cm \epsfbox{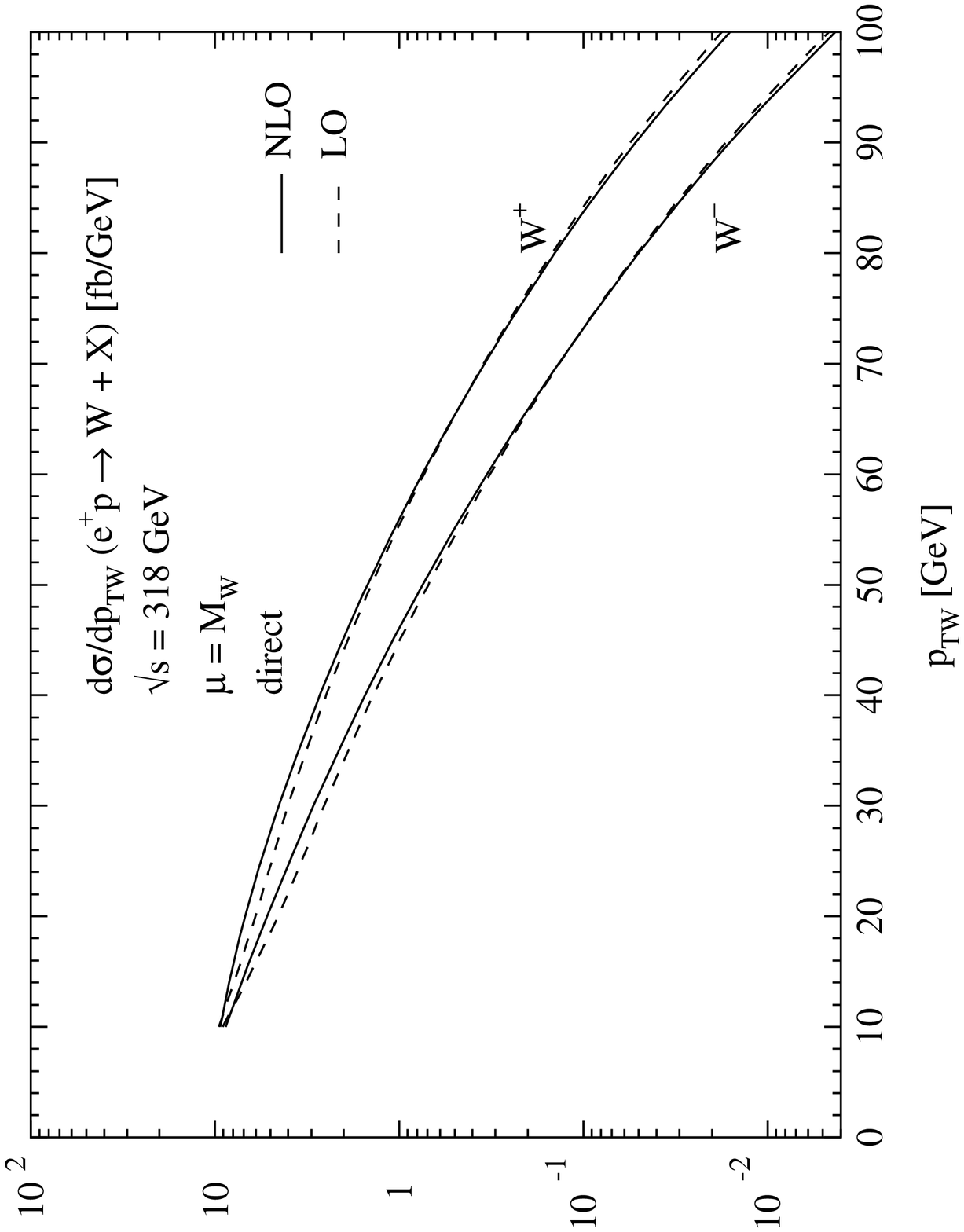}
\end{turn}
\vspace*{0.0cm}

\vspace*{0.5cm}
\begin{turn}{-90}%
\epsfxsize=8cm \epsfbox{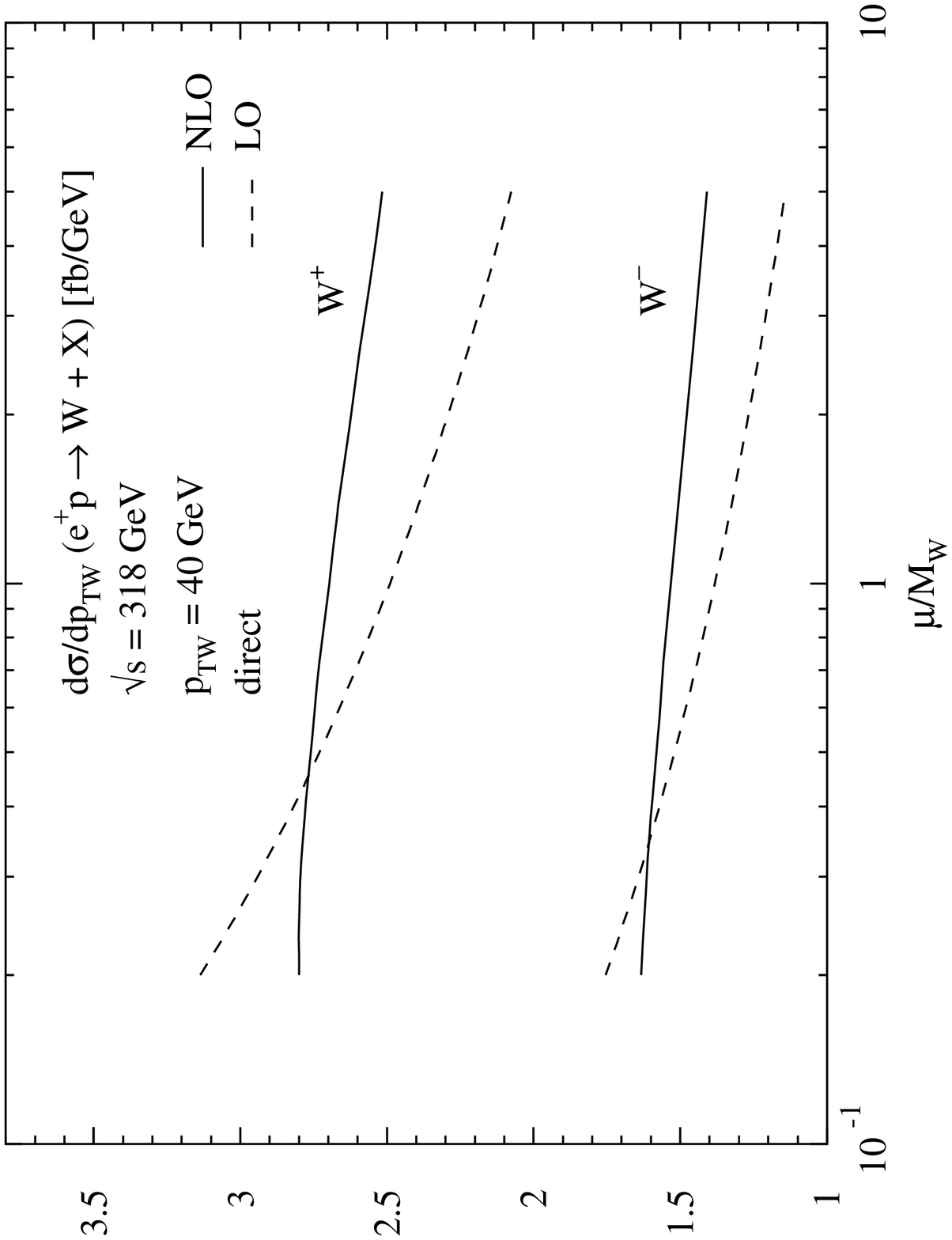}
\end{turn}
\vspace*{-0.0cm}

\begin{picture}(20,154)
\put(60,445){\sf\LARGE (a)}
\put(60,202){\sf\LARGE (b)}
\end{picture}

\vspace*{-5.5cm}

\caption[]{\it \label{fg:pti} (a) Transverse momentum distribution of $W$
bosons at HERA for direct photoproduction. (b) Dependence of the direct
contribution 
to $W$ production on the
renormalization and factorization scale $\mu = \mu_F = \mu_R = \xi M_W$
for $p_{TW}=40$ GeV.
The full curves represent the NLO and the dashed curves the
LO predictions.}
\end{figure}

Setting $\mu_R = \mu_F = M_W$ for the
renormalization and factorization scales, we present the final results for
the $p_{TW}$ distributions of $W+X$ production in
Fig.~\ref{fg:pti} (a).
The QCD corrections modify the direct contribution by about $\pm (10-15)\%$ and
are thus of moderate size.
To estimate the theoretical uncertainties, the
renormalization/factorization scale dependence of the direct
contributions to the processes $e^+ p \to W^\pm + X$
is presented in Fig.~\ref{fg:pti} (b) for HERA conditions. The scale
dependence is significantly smaller, once the NLO corrections are
included. The residual scale dependence is reduced from about 20\% down
to about 5\%. Fig.\,\ref{fg:pti} (b) clearly indicates that the NLO QCD
corrections are accidentally small at the central scale determined by
the $W$ boson mass.
Since the uncertainties of the parton densities are of similar size, the
total theoretical uncertainty can be estimated to be less than about 10\%.

\section{Conclusions}
We have presented predictions for $W$ boson production at HERA including
the QCD corrections to the dominant direct photon mechanism at finite
transverse momentum of the $W$ boson. We find that the QCD corrections are
of moderate size. 
However, the QCD corrections to the DIS part are still
unknown. They are not expected to be significantly larger, because they
have to cancel the $Q^2_{max}$ dependence of the NLO direct contribution
and thus have to be of similar size.
Since the QCD corrections are dominated by soft gluon
effects, the shapes of the differential distributions are hardly
affected. Therefore the results obtained in this work cannot explain the
excess of isolated high-energy lepton events observed at the H1
experiment \cite{muon}. \\

\noindent {\bf Acknowledgements.}\\
I would like to thank K.P.~Diener and M.~Spira for the fruitful collaboration.

\end{document}